\documentclass[aps,prb,twocolumn, 
superscriptaddress]{revtex4}
\usepackage{tabularx}
\usepackage{bm}
\usepackage{euscript}
\usepackage{epsfig,psfrag,subfigure}
\usepackage{graphicx}
\usepackage{color}
\usepackage{amsfonts}
\usepackage{exscale}
\usepackage{amsbsy}
\usepackage{dcolumn}

\begin{document}

\title{ Spin-Valley Kondo Effect in Multi-electron Silicon Quantum
Dots }
\author{Shiue-yuan Shiau, and Robert Joynt}
\affiliation{Department of Physics, University of Wisconsin, 1150 University Avenue,
Madison, Wisconsin 53706}
\date{\today}

\begin{abstract}
We study the spin-valley Kondo effect of a silicon quantum dot occupied by $%
\mathcal{N}$ electrons, with $\mathcal{N}$ up to four. We show that the
Kondo resonance appears in the $\mathcal{N}=1,2,3$ Coulomb blockade regimes,
but not in the $\mathcal{N}=4$ one, in contrast to the spin-1/2 Kondo
effect, which only occurs at $\mathcal{N}=$ odd. Assuming large orbital
level spacings, the energy states of the dot can be simply characterized by
fourfold spin-valley degrees of freedom. The density of states (DOS) is
obtained as a function of temperature and applied magnetic field using a
finite-$U$ equation-of-motion approach. The structure in the DOS can be
detected in transport experiments. The Kondo resonance is split by the
Zeeman splitting and valley splitting for double- and triple-electron Si
dots, in a similar fashion to single-electron ones. The peak structure and
splitting patterns are much richer for the spin-valley Kondo effect than for
the pure spin Kondo effect.
\end{abstract}

\maketitle

\section{Introduction}

Single- and few-electron quantum dots (QDs) coupled to leads have been
realized in the last several years \cite{Ciorga2000}. \ Their theoretical
description is similar to that of a magnetic impurity in a bulk metal, a
problem that has been studied for decades. \ It gives rise, among other
things, to Kondo physics \cite{Hewson}. \ The advantage of the QD system is
that many parameters can be varied by adjusting voltages on the electrodes
that surround the QD, whereas in the bulk these parameters are fixed \cite%
{Folk1996}. \ One of these parameters is the total number of electrons $%
\mathcal{N}$ on the QD. \ Of course in many cases some of the electrons on
the QD can be considered as "core" electrons. \ This works only when the
Coulomb interaction or the dot orbital level spacings are large and
therefore all but one or two of the electrons residing on the closest
orbital to the Fermi energy on the dot can be treated as vacuum. In this
single-electron dot picture the spin-1/2 Kondo effect ensues from spin
fluctuation of the dot electron spin coupled to the conduction electrons on
the leads. As a result, the dot electron spin binds to an electron spin on
the leads to form a spin singlet. \ When temperatures and magnetic field
splittings are lower than the energy scale characteristic of this spin
singlet, a narrow zero-bias Kondo resonance crops up in the differential
conductance as a function of source-drain voltage. \ 

For GaAs QDs \cite{Cronenwett1998,Goldhaber1998}, the Kondo effect is only
observed in $\mathcal{N}=$ odd Coulomb blockade regimes. \ This is the only
case in which it is possible for a spin singlet formed by the highest-energy
electron on the dot and electrons in the leads to be the ground state, and
then only provided that finite orbital spacings render it energetically
favorable. \ No Kondo resonance is formed in $\mathcal{N}=$ even Coulomb
blockade regimes, since the highest-energy dot electrons themselves will
normally form a spin singlet. \ Thus a characteristic signature of the Kondo
effect has always been this periodicity of 2 in the variable $\mathcal{N}.$
\ A second characteristic signature is the magnetic field dependence. \ The
Kondo resonance, which manifests itself as a peak at zero voltage when the
conductance is measured as a function of source-drain voltage, splits into
two peaks due to the Zeeman effect. \ These peaks are separated by $\delta
V=2g\mu _{B}B.$

For Si QDs, experiments on the Kondo effect have so far been able to reach
the stage of few-electron dots, but not that of single-electron ones. In a
few-electron Si dot \cite{Levente2007}, a Kondo resonance at zero bias
voltage is split into two peaks by an applied magnetic field. The value of
the extracted $g$ factor is about 2.26, slightly larger than the expected 2,
suggesting a possible contribution from the valley degree of freedom. \ No
distinct periodicity has yet been observed.

We note one apparent exception to the periodicity rule in GaAs. \ A recent
report \cite{Sasaki2000} on the spin-orbital Kondo effect in an integer-spin
QD has shown a Kondo effect in an $\mathcal{N}=$ even Coulomb blockade
regimes. Considering two orbital levels on the dot, the condition of a spin
singlet-triplet degeneracy can be artificially achieved by tuning the
magnetic field and a Kondo resonance emerges at a particular magnitude of
field. When further increasing the field, this resonance splits into two
peaks at a finite bias voltage, and the separation of the two peaks is twice
the singlet-triplet energy difference. In perpendicular field, their
difference has a much stronger field dependence than the Zeeman splitting.
The smaller-scale Zeeman splitting was not observed in the data. The spin
singlet-triplet Kondo effect involving two orbital levels in $\mathcal{N}=$
even regimes resembles in some respects the spin-valley Kondo effect in
even-electron Si dots, the main body of this study, in the sense that the
degeneracy does not come entirely from spin.

Silicon is an indirect bandgap semiconductor. \ The top of the valence band
lies at the $\Gamma $-point, while the conduction band has six degenerate
minima along the $\Gamma $-$X$ directions. \ The conduction electrons in\ $n$%
-type silicon have therefore a sixfold degeneracy corresponding to this
valley degree of freedom. \ Si QDs are formed in heterostructures where the
active layer is a thin layer of pure silicon sandwiched between layers of Si$%
_{1-x}$Ge$_{x}$ alloy, which puts the silicon layer in a state of in-plane
tensile strain. \ This raises four of the conduction band minima by a large
energy ($\sim 0.1$ eV), leaving only a twofold valley degeneracy. \ This in
turn is split by small effects that break the mirror symmetry (reflection
through the $x$-$y$ plane). \ This small ($<$ 1 meV) splitting is enhanced
by a perpendicular magnetic field. \ It can be controlled by changing
electrostatic and magnetic confinement \cite{Goswami2007}. On the other
hand, a two-dimensional tight-binding model \cite{Shiau2007} predicts the
possibility of valley index nonconservation during tunneling from the leads
to the dots, which results in opening up additional tunneling channels
between an even (odd) valley state on the leads and an odd (even) valley
state on the dot, which changes significantly the features of the Kondo
resonance in single-electron Si QDs. The valley mixing is made possible by
the rough interfaces that confine the two dimensional electron gas (2DEG) of
strained Si, in which the dot-leads system is formed. \ 

This valley near-degeneracy is a potential source of leakage and decoherence
in quantum computing schemes in which Si QDs serve as the qubits. \ This is
an additional reason for trying to understand its consequences. \ 

In earlier work, we showed that valley degeneracy produces a novel Kondo
effect in $\mathcal{N}=$ 1 Si QDs \cite{Shiau2007} . We will show below that
for double-electron Si dots, there is also a Kondo resonance, which suffers
both the valley and Zeeman splittings, in contrast to the Kondo resonance
affected by the field-dependent singlet-triplet energy difference mentioned
above \cite{Sasaki2000}.

Figure 1 in Ref.~\onlinecite{Shiau2007} characterizes the four spin-valley
energy levels as a function of magnetic field for a single orbital in a Si
QD. This energy-level structure draws on two experiments\cite{Goswami2007}
that observed a valley splitting in Hall bars and Quantum Point Contacts
(QPCs); in the first the valley splitting shows a linear dependence with
applied field, in the latter the magnitude is measured to be about
1 meV. Furthermore, a finite zero-field valley splitting was observed in
both experiments, in which at zero field the splitting in Hall bars is about 
$1.5\pm 0.6~\mu $eV, much smaller than in QPCs . The difference is ascribed
to interface disorder \cite{Friesen2006,Kharche2007}. This level structure
will again be used later to define the dot energy levels in Sec.~\ref%
{sect:DOS}.

\begin{figure}[tph]
\vspace{-6 cm} \centering      
\subfigure[\label{fig:config1a}]  {
\includegraphics[width=0.45\textwidth,clip]{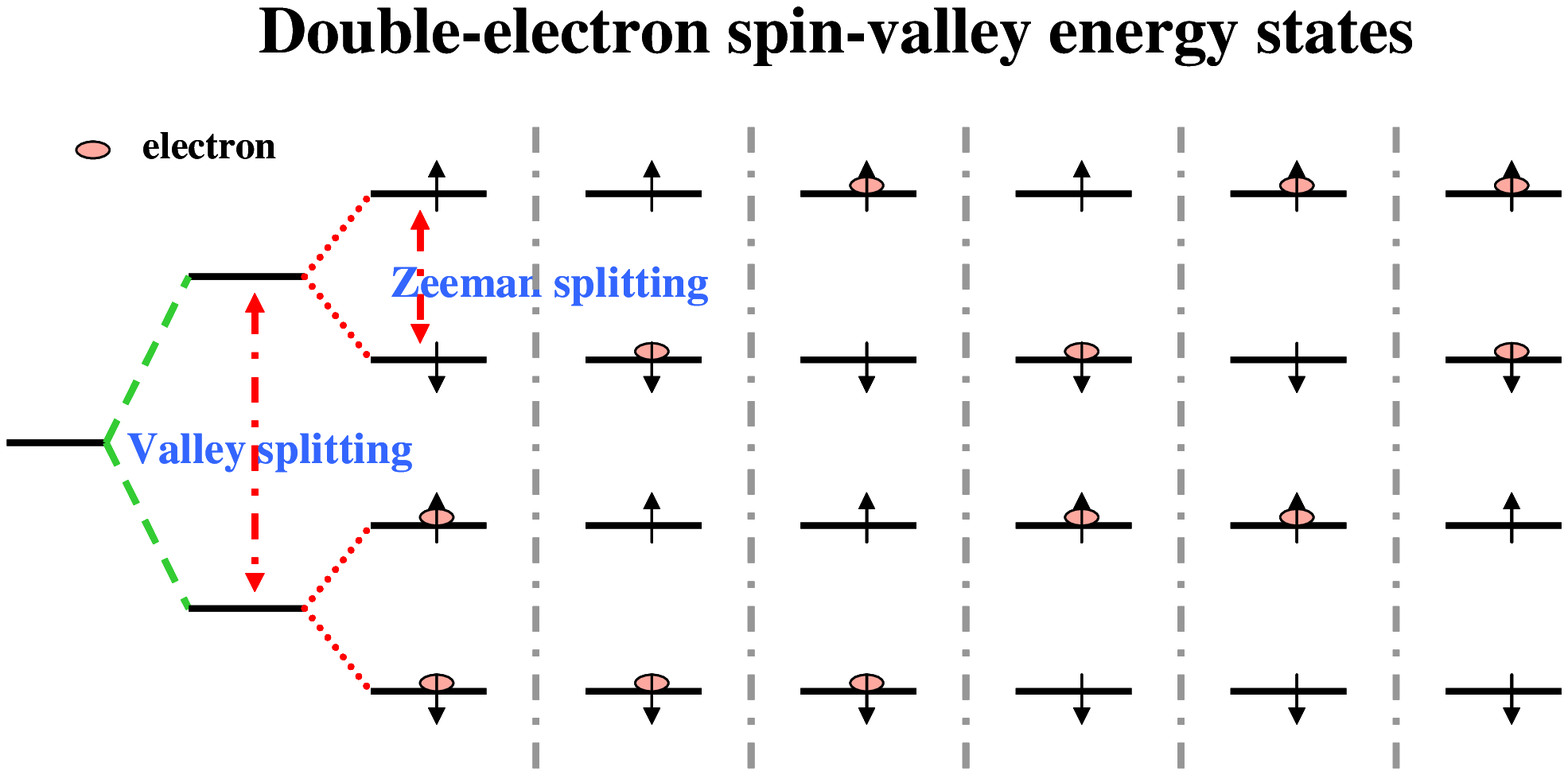}} \\  
\vspace{-2.5 in} 
\subfigure[\label{fig:config1b}]  { \hspace{1 cm}
\includegraphics[width=0.45\textwidth,clip ]{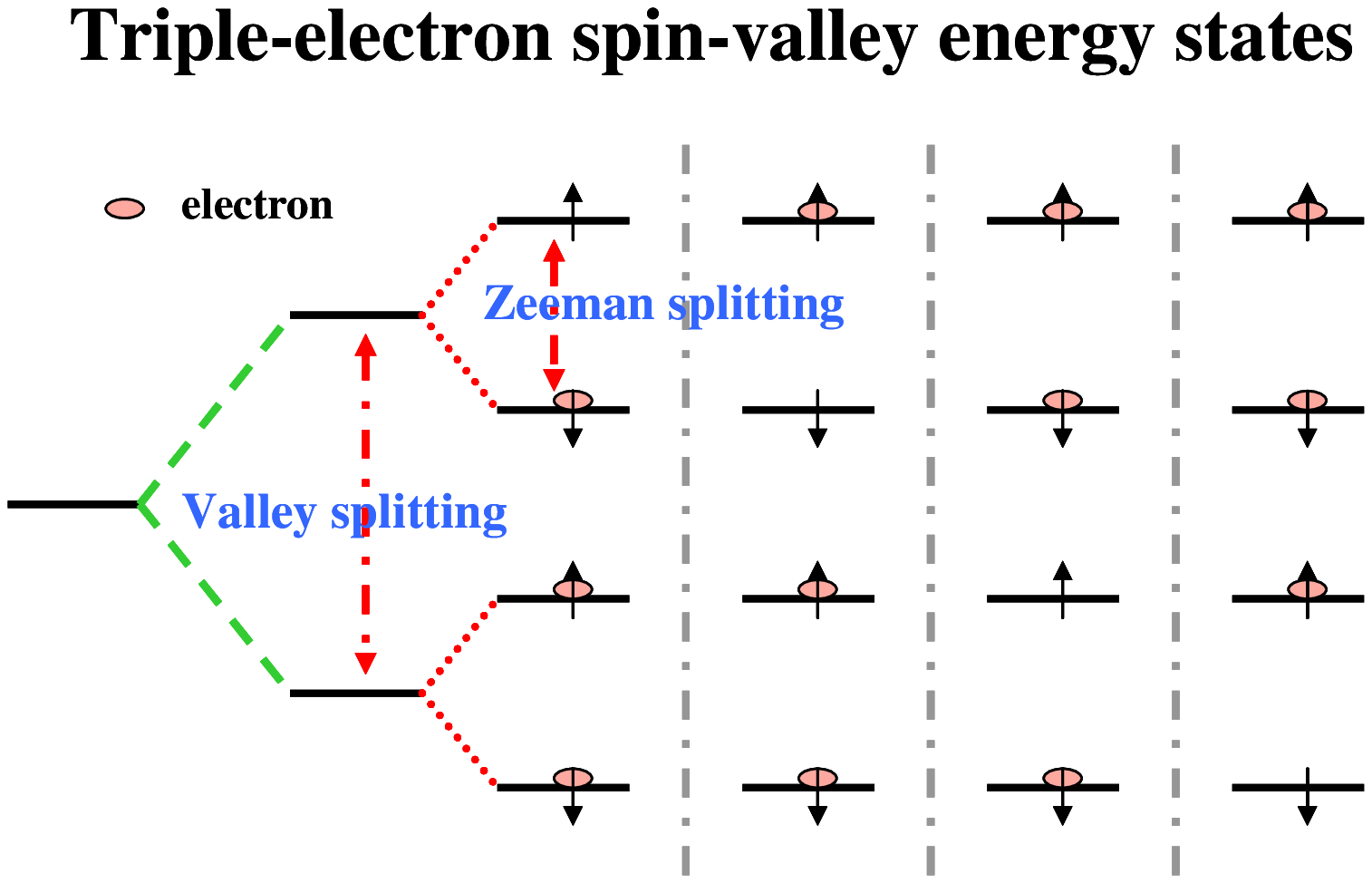}} 
\caption{{\protect\small {When the Coulomb interaction is large, the
electron occupation number $\mathcal{N}$ is a good quantum number at weak
couplings. Shown here are the six and four configurations of low-energy
spin-valley dot states for the (a) $\mathcal{N}=2$ and (b) $\mathcal{N}=3$
regimes, respectively. The spin-valley Kondo effect is related to the $%
\mathcal{N} $-conserved many-body transitions where these dot states
interact with the lead states. Note that the relative size of the valley and
Zeeman splittings in the cartoon figures does not necessarily reflect every
tested sample since the valley splitting is sample-dependent.}}}
\label{fig:config1}
\end{figure}

Our aim in this paper is to describe the Kondo resonance(s) in Si QDs for $%
\mathcal{N}>1.$ \ The Kondo effect itself results from a ground state in a
which the dot spin and valley states are mixed with the lead states to form
a singlet ground state. \ The resonance in transport occurs because the
resulting wave function has weight at or near the Fermi energies of both
leads, leading to a zero-bias or near-zero-bias anomaly. \ The anomaly can
be shifted and split by a magnetic field. \ For $\mathcal{N}=$ 1, we showed
that the zero-bias resonance is split by both the valley and Zeeman
splittings so that there are more split peaks in non-linear $I-V$
characteristics than in the spin-1/2 Kondo effect \cite{Shiau2007}. To
complete the whole picture, we continue to investigate whether there is a
spin-valley Kondo effect in the $\mathcal{N}=2,3,4$ Coulomb blockade regimes
and if so, how the splittings occur in each regime. We use again an
equation-of-motion approach to obtain the interacting DOS that will be shown
below to directly reflect the differential conductance. We assume orbital
level spacings to be larger than a few times the dot Coulomb interaction $U$%
, so that the orbitals are well separated in energy. As a result, it is
sufficient to only consider the single orbital closest to the Fermi energy.
This condition could be satisfied in a small dot size of $\sim $10 nm. Then,
since the Coulomb interaction is large, the charge fluctuation is small
enough for the electron occupation number $\mathcal{N}$ to be a well-defined
quantum number between the Coulomb blockade peaks. \ Then what needs be
taken into account is the low-energy electron configurations for each $%
\mathcal{N}$. Figure \ref{fig:config1} demonstrates the configurations of
these spin-valley energy states for $\mathcal{N}=2,3$ \ that interact with
the lead states.

In the next section we describe the model and our calculation method. \ The
results for the Kondo temperature follow in the section after that, and then
we plot and discuss the DOS for various cases. \ Finally we summarize and
give some conclusions.

\section{Equation-of-motion approach}

The equation-of-motion (EOM) approach has proven to be a good tool for
investigating the density of states (DOS) in the Kondo effect \cite{Czycholl}%
. \ Since this quantity is the one that interests here, we shall employ this
method. \ In particular it is able to handle arbitrary orbital structure and
particle number, as well as finite temperature and the presence of a
magnetic field. \ The basic technical details are given in Ref. %
\onlinecite{Shiau2007}, so here we describe only those additional features
that are required to treat the $\mathcal{N}>1$\ case.

A Hamiltonian that describes a system consisting of the single-particle
energy levels of the leads, the dot, the tunneling matrix couplings that
connect the levels of the leads and the dot, as well as the Coulomb
interaction between electrons on the dot is the Anderson impurity model,
expressed by 
\begin{widetext}
 \begin{eqnarray}
 \mathcal{H}&=&\sum_{ik,m\sigma} \varepsilon_k c_{ikm\sigma}^{\dag} c_{ikm\sigma}+\sum_{m\sigma}\varepsilon_{m\sigma}f^\dag_{m\sigma} f_{m\sigma} +
\sum_{ikm\sigma } V_{O,ik}(c^\dag_{ikm\sigma}f_{m\sigma} + f_{m\sigma}^\dag c_{ikm\sigma}) +\sum_{ikm\sigma }V_{X,ik}(c_{ikm\sigma }^
{\dag}f_{\bar{m}\sigma }+f_{\bar{m}\sigma%
}^{\dag}c_{ikm\sigma})\nonumber \\
 && +\frac{U}{2}\sum_{m^\prime \sigma^\prime \not= m\sigma}n_{m^\prime\sigma^\prime}n_{m\sigma}\nonumber
 \end{eqnarray}
 \end{widetext}Here the spin-1/2 index $\sigma \in \left\{ \uparrow
,\downarrow \right\} $, and the valley index $m\in \left\{ \text{%
e(even),~o(odd)}\right\} $. Even and odd denoted the two valley states. $\ 
\bar{m}$ is the opposite of the valley index $m$. The operator $c_{ikm\sigma
}^{\dag }(c_{ikm\sigma })$ creates (annihilates) an electron with an energy $%
\varepsilon _{k}$ in the $i$ lead, $i\in L,R$, while the operator $%
f_{m\sigma }^{\dag }(f_{m\sigma })$ creates (annihilates) an electron with
an energy $\varepsilon _{m\sigma }$ on the QD, connected to the leads by
Hamiltonian intravalley coupling $V_{O,ik}$ and intervalley coupling $%
V_{X,ik}$. \ For perfect interfaces at the boundaries of the well, it could
happen that $V_{X,ik}=0,$ i.e., that the valley index is conserved in
tunneling. \ For real dots, we have shown that $V_{O,ik}$ and $V_{X,ik}$ are
likely to be the same order of magnitude \cite{Shiau2007}. \ We assume that $%
V_{O(X)ik}$ does not depend on the spin index $\sigma $. $U$ is the Coulomb
interaction on the dot and is assumed to be independent of the valley
index.\ 

The Kondo effect can be observed by measuring the current $I$ and likewise
the differential conductance $G$ as a function of source-drain voltage $%
V_{sd}.$ \ Theoretically the differential conductance $G=dI/dV_{sd}$ is
given by differentiating the generalized Landauer formula, given in Ref.~%
\onlinecite{Shiau2007}. The differential conductance is approximately
proportional to the interacting DOS --- DOS $=-\mathrm{Im}[\mathcal{G}%
_{m\sigma }(eV_{sd})]/\pi $ --- given the assumption of an initially flat
noninteracting DOS in the leads. Although the lead DOS and tunneling matrix
elements vary with applied voltages, it is usually true that the variations
are slow compared with the sharp Kondo resonance structures. \ When this is
true, understanding the dot DOS is sufficient to identify the fine structure
in the conductance near zero bias.\

Thus we need to compute $\mathcal{G}_{m\sigma }(w),$ the retarded Green's
function: 
\begin{eqnarray}
\mathcal{G}_{m\sigma }(w) &\equiv &\langle \langle f_{m\sigma },f_{m\sigma
}^{\dag }\rangle \rangle  \nonumber \\
&=&-i\int_{0}^{\infty }e^{i\alpha t}\langle \{f_{m\sigma }(t),f_{m\sigma
}^{\dag }(0)\}\rangle dt.  \nonumber
\end{eqnarray}%
where $\alpha =w+i\delta $. We compute the equation of motion for $\mathcal{G%
}_{m\sigma }(w)$ in the frequency domain: 
\begin{eqnarray}
w\langle \langle A,B\rangle \rangle &=&\langle \{A,B\}\rangle +\langle
\langle \lbrack A,H],B\rangle \rangle  \nonumber \\
&=&\langle \{A,B\}\rangle +\langle \langle A,[H,B]\rangle \rangle . 
\nonumber
\end{eqnarray}%
By applying the above equation of motion to a Green's function, we obtain
higher-order Green's functions on the right-hand side of the equation, which
we further expand by repeating the same procedure until all second-order ($%
\mathcal{O}(V^{2})$) contributions are preserved after the decoupling scheme%
\cite{Czycholl}. After some tedious but straightforward calculations, we
acquire equations of motion that couple the Green's functions $\mathcal{G}%
_{m\sigma }(w),$ $\langle \langle f_{\bar{m}\sigma },f_{m\sigma }^{\dag
}\rangle \rangle ,$ $\langle \langle n_{l}f_{m\sigma },f_{m\sigma }^{\dag
}\rangle \rangle ,$ $\langle \langle n_{l^{\prime }}f_{\bar{m}\sigma
},f_{m\sigma }^{\dag }\rangle \rangle ,$ $\langle \langle
n_{l}n_{j}f_{m\sigma },f_{m\sigma }^{\dag }\rangle \rangle ,$ $\langle
\langle n_{l^{\prime }}n_{j^{\prime }}f_{\bar{m}\sigma },f_{m\sigma }^{\dag
}\rangle \rangle ,$ $\langle \langle n_{l}n_{j}n_{p}f_{m\sigma },f_{m\sigma
}^{\dag }\rangle \rangle ,$ $\langle \langle n_{l^{\prime }}n_{j^{\prime
}}n_{p^{\prime }}f_{\bar{m}\sigma },f_{m\sigma }^{\dag }\rangle \rangle $ ($%
\{l,j,p\}\not=m\sigma $ and $\{l^{\prime },j^{\prime },p^{\prime }\}\not=%
\bar{m}\sigma $ are shorthand notations for both $m$ and $\sigma $ indices)
that describe single, double, triple and quadruple occupancies,
respectively. These Green's functions suffice to describe fourfold
degenerate Si QDs that can host up to four electrons. They are nothing but
linearly coupled matrix elements spanned by spin and valley quantum numbers,
and they can be solved for by linear diagonalization in terms of their
coefficients: second-order perturbation terms, integral functions,
occupation numbers $\langle n_{m\sigma }\rangle $, $\langle n_{l}n_{m\sigma
}\rangle ,$ $\langle n_{l}n_{j}n_{m\sigma }\rangle ,$ and expectation values 
$\langle f_{m\sigma }^{\dag }f_{\bar{m}\sigma }\rangle ,$ $\langle
n_{l^{\prime \prime }}f_{m\sigma }^{\dag }f_{\bar{m}\sigma }\rangle ,$ $%
\langle n_{l^{\prime \prime }}n_{j^{\prime \prime }}f_{m\sigma }^{\dag }f_{%
\bar{m}\sigma }\rangle $ ($l^{\prime \prime },j^{\prime \prime }\not=m\sigma
,\bar{m}\sigma $) . It is noteworthy that some perturbation terms and the
integral functions are logarithmically divergent at the Fermi energy,
thereby giving rise to a zero-bias anomaly, and have to be treated carefully
in the Kondo regime $T\leq T_{K}$ ($T_{K}$ is the Kondo temperature that we
will define later). Details are parallel to those in Ref.~%
\onlinecite{Shiau2007}, so that we do not give them here, except to note the
following:\

First, we assume a flat and symmetric noninteracting DOS in the source and
drain, so $V_{O(X),ik}=V_{O(X)}$. Since the valley index is not conserved,
it is convenient to introduce $V_{O}=V\cos {\phi },~V_{X}=V\sin {\phi }$. We
define $V^{2}=V_{O}^{2}+V_{X}^{2}$ and $2V_{O}V_{X}=\beta V^{2}$ with $\beta
=\sin {2\phi }$ and $0\leq \beta \leq 1$. Parameter $\beta $ gauges the
extent of valley index nonconservation. In other words, $\beta =0$ implies
valley index conservation, reflecting perfectly smooth interfaces of the
2DEG, whereas $\beta =1$ the maximal valley mixing.

To compute numerical results, we use the following integral 
\[
\int^D_{-D} dw^{\prime }\frac{f_{FD}(w^{\prime })}{w-w^{\prime }\pm i\delta }%
=-\Psi (\frac{1}{2}\pm \frac{w-\varepsilon_F}{2\pi i T}) +\text{ln}\frac{D+w%
}{2\pi T} \mp \frac{i\pi}{2} 
\]
where the parameter $D$ is the conduction half-bandwidth, and $f_{FD}$ the
Fermi function. $\Psi(z)$ is the digamma function that asymptotically
behaves as $\text{ln}z$ as $|z|\gg 1$. This logarithmic divergence produces
a low energy scale in the Kondo regime, the domain of our interest. This
scale is defined as the Kondo temperature. 
We also assume the self-energy term 
\[
\Sigma_0(w)=\sum_{ik} \frac{V^2}{w-\varepsilon_k+i\delta}\simeq -i \Gamma 
\]
where $\Gamma=\pi V^2/D$. This approximation is valid near the Fermi energy
which is the region of most experimental interest. \ 

The integral functions come from the correlation functions of the dot and
leads after decoupling the higher-order Green's functions. By simple
transformation these correlation functions can be rewritten in terms of
integral functions over the Green's functions shown above, and the set of
equations of motion terminates after the decoupling \cite{Shiau2007}. To
compute the integral functions whose integrands contain the Green's
functions, first we use the approximation adopted by Lacroix\cite%
{Lacroix1981} and V. Kashcheyevs \textit{et al.} \cite{Kashcheyevs2006} who
assume that the integral functions are dominated by the singularity and only
strongly affect the region around the Fermi energy. As a result , we can
approximate, for instance, the integral function 
\begin{equation}
\Gamma \int dw^{\prime }\frac{f_{FD}(w^{\prime })\mathcal{G}_{m\sigma
}^{\ast }(w^{\prime })}{w-w^{\prime }}\simeq \Gamma \mathcal{G}_{m\sigma
}^{\ast }(\varepsilon _{F})\int dw^{\prime }\frac{f_{FD}(w^{\prime })}{%
w-w^{\prime }}  \label{eq:integral1}
\end{equation}%
Likewise for other integral functions.

This leads to a set of coupled integral equations for the exact Green's
functions since they are functions of integrals of themselves. Our strategy
here is to make a guess of the expected structures of the Green's functions,
substitute them for the exact ones in all the integral functions, and
iterate to self-consistency. To illustrate, we replace the Green's function $%
\mathcal{G}_{m\sigma }(w)$ in Eq.~\ref{eq:integral1} by $G_{m\sigma
}^{(at)}(w)$ in Ref.\onlinecite{Yeyati_1999} as the expected Green's
function. If we assume $\Gamma /U\ll 1$, we find $\mathcal{G}_{m\sigma }(w)$
consistent with the features of $G_{m\sigma }^{(at)}(w)$, and equivalent to $%
G_{m\sigma }^{(at)}(w)$ at high temperatures where the integral functions
are small and can be disregarded. Moreover, to mimic the Green's functions
that feature a broad peak of width $\sim \Gamma $ centered around the
discrete bare dot energy levels, each propagator $[w-\varepsilon _{m\sigma
}-dU]^{-1}$ in $G_{m\sigma }^{(at)}(w)$ is given a finite spectral width $%
\Gamma _{d}$ where $d=0,1,2,3$. $\Gamma _{d}$'s take into account only the
self-energy terms $\Sigma _{0}(w)$'s in each propagator. Thus we assign 
\[
\Gamma _{0}=\Gamma ,~~\Gamma _{1}=3\Gamma ,~~\Gamma _{2}=5\Gamma ,~~\Gamma
_{3}=7\Gamma . 
\]%
This increase of spectral widths mimics real physical systems where the
potential barrier gradually opens up at high energy. Despite the fact that $%
\Gamma _{d}$'s are not real spectral widths, the errors contribute only $%
\mathcal{O}(\Gamma ^{2})$ to the integral functions that already have a
prefactor $\Gamma $(c.f. Eq.~\ref{eq:integral1}).\ 

We shall operate under the assumption that $\Gamma /U\ll 1$, yet we are
interested in the Kondo regime. Hence this approximation for the integral
functions might not be valid. However, as argued by Czycholl\cite{Czycholl},
the approximated integral functions should not deviate very much from the
true ones, so long as the temperature is not too far below the Kondo
temperature and higher than second-order contributions can be legitimately
neglected.

On the other hand the occupation numbers and other expectation values are
computed by integration over the Green's functions. For instance 
\begin{eqnarray}
\langle n_{m\sigma }\rangle &=&-\frac{1}{\pi }\int dwf_{FD}(w)\mathrm{Im}%
\mathcal{G}_{m\sigma }(w).  \nonumber \\
\langle n_{l}n_{m\sigma }\rangle &=&-\frac{1}{\pi }\int dwf_{FD}(w)\mathrm{Im%
}\langle \langle n_{l}f_{m\sigma },f_{m\sigma }^{\dag }\rangle \rangle . 
\nonumber \\
\langle f_{m\sigma }^{\dag }f_{\bar{m}\sigma }\rangle &=&-\frac{1}{\pi }\int
dwf_{FD}(w)\mathrm{Im}\langle \langle f_{\bar{m}\sigma },f_{m\sigma }^{\dag
}\rangle \rangle .  \nonumber \\
\langle n_{l^{\prime \prime }}f_{m\sigma }^{\dag }f_{\bar{m}\sigma }\rangle
&=&-\frac{1}{\pi }\int dwf_{FD}(w)\mathrm{Im}\langle \langle n_{l^{\prime
\prime }}f_{\bar{m}\sigma },f_{m\sigma }^{\dag }\rangle \rangle .  \nonumber
\end{eqnarray}%
Now that all the perturbation terms, integral functions, the occupation
numbers, and other expectation values in the Green's functions have been
accounted for, we proceed to iterate to self-consistency.\ 

In Sec.~\ref{secKondoT}, we show the Kondo temperatures for $\mathcal{N}=1,2,3$
regimes as a function of finite Coulomb interaction $U$, dot energy level $%
\varepsilon _{m\sigma }$, coupling strength $\Gamma $, and parameter $\beta $.%
 \ We also present numerical results of the DOS as a function of energy, in which a
Kondo resonance is found at zero bias voltage and splits when applying a
magnetic field, as demonstrated in Sec.~\ref{sect:DOS}.

\section{The $\protect\beta$-dependent Kondo temperatures\label{secKondoT}}

It is enlightening to start first with a simple case: degenerate spin-valley
states. Although it is unlikely to observe a degenerate spin-valley Kondo
effect in Si dots since the valley splitting is not zero even in the absence
of external field, it is still useful for us to first derive the Kondo
temperatures from the solutions near the Fermi energy of the real parts of
the denominators in $\mathcal{G}_{m\sigma }(w)$\cite{Hewson}. Assuming $%
D>|\varepsilon _{m\sigma}|,U\gg T_{K}$'s and at zero temperature, we obtain 
\begin{widetext}
\begin{eqnarray}
T_{K1}^{\pm }(\beta ) &\simeq &D^{\ast }\exp{\frac{\pi (\varepsilon
_{m\sigma}-\varepsilon _{F})(\varepsilon _{m\sigma}+U-\varepsilon _{F})}{%
(3\pm \beta )\Gamma U}}  \nonumber \\
T_{K2}^{\pm }(\beta ) &\simeq &D^{\ast }\exp{\frac{\pi (\varepsilon
_{m\sigma}-\varepsilon _{F})(\varepsilon _{m\sigma}+U-\varepsilon
_{F})(\varepsilon _{m\sigma}+2U-\varepsilon _{F})}{(3\pm \beta )\Gamma
U[4/3(\varepsilon _{m\sigma}-\varepsilon _{F})+(\varepsilon
_{m\sigma}+2U-\varepsilon _{F})]}}  \nonumber \\
T_{K3}^{\pm }(\beta ) &\simeq &D^{\ast }\exp{\frac{\pi (\varepsilon
_{m\sigma}+U-\varepsilon _{F})(\varepsilon _{m\sigma}+2U-\varepsilon
_{F})(\varepsilon _{m\sigma}+3U-\varepsilon _{F})}{(3\pm \beta )\Gamma
U[(\varepsilon _{m\sigma}+U-\varepsilon _{F})+4/3(\varepsilon
_{m\sigma}+3U-\varepsilon _{F})]}}  \nonumber
\end{eqnarray}%
\end{widetext}
Where $D^{\ast }\equiv D|2\varepsilon _{m\sigma}+U-\varepsilon _{F}| / %
| D+2\varepsilon _{m\sigma}+U| $. $T_{K1}^{\pm }(\beta ),~T_{K2}^{\pm
}(\beta ),~T_{K3}^{\pm }(\beta )$ are the $\beta $-dependent Kondo
temperatures for single, double, and triple occupancies, respectively. We
have already shown\cite{Shiau2007} that there are two Kondo
temperatures---equivalent to $T_{K1}^{\pm }(\beta )$ in the infinite $U$
limit--- in single-electron Si dots if the valley index is not conserved ($%
\beta >0$), and the parameter $\beta $ greatly enhances one Kondo
temperature while suppressing the other in a similar fashion. As a result, $%
T_{K1}^{+}(\beta )$, the largest of the two, dominates the screening of the
dots. Here we show the same with double-electron and triple-electron Si
dots. But when $\varepsilon _{m\sigma}+3U<\varepsilon _{F}$, the spin-valley
Kondo effect disappears since all energy levels are occupied and inelastic
transitions are prohibited. A keen observation will give that these $%
T_{K}^{\pm }(\beta )$'s are not unrelated. Indeed, they obey a simple
relation 
\begin{equation}
\frac{1}{\mathrm{ln}T_{K2}^{\pm }(\beta )/D^{\ast }}\simeq \frac{1}{\mathrm{%
ln}T_{K1}^{\pm }(\beta )/D^{\ast }}+\frac{1}{\mathrm{ln}T_{K3}^{\pm }(\beta
)/D^{\ast }}  \label{Tk_log}
\end{equation}%
In other words, the Kondo temperatures in the $\mathcal{N}$ regime are
influenced by the ones in the neighboring $\mathcal{N}-1$ and $\mathcal{N}+1$
regimes. A similar logarithmic relation was found in Ref.~%
\onlinecite{Pustilnik2000}, where although a Kondo effect is produced in $%
\mathcal{N}=$ even regimes, it belongs to a two-level QD. Consequently it is
generically different from a fourfold degenerate Si QD.

The larger Kondo temperatures $T_{K}^{+}(\beta )$'s express the energy
scales where the spin-valley Kondo effect can be observed in the Kondo
regime, namely, at temperatures lower than $T_{K}^{+}(\beta )$'s. The EOM
approach produces a finite-$U$ Kondo temperature $T_{K1}^{+}(\beta =0)$ that
is different by a prefactor $3/4$ in the exponent from the one using a
scaling theory\cite{Eto2006}. The discrepancy of this approach was claimed
to be due to neglect of higher-order corrections\cite{Luo1999}. It should
therefore be stressed that $T_{K}^{\pm }(\beta )$'s may only show a
qualitative dependence on the model parameters. Despite this defect, $%
T_{K1}^{\pm}(\beta )$'s computed from the EOM approach scale logarithmically
the zero-bias resonance in single-electron Si QDs\cite{Shiau2007}, thereby
portraying a Kondo-like effect, as will be expected the same with $%
T_{K2}^{\pm}(\beta )$ and $T_{K3}^{\pm}(\beta )$. The larger $%
T_{K}^{+}(\beta )$'s will serve as energy scales for the demonstration of
the spin-valley Kondo resonances in the next section.

It is noteworthy that at zero temperature the above solutions exist except
when $\varepsilon _{m\sigma }=(\varepsilon _{F}-U)/2$. This particular
condition leads to a particle-hole symmetry. The EOM approach fails to
produce at this point an energy scale that governs the low temperature
behavior. This scenario requires a special mathematical treatment \cite%
{Kashcheyevs2006,Appelbaum1969}. Regardless of its intrinsic interest, this
case is not our main concern in this paper.

\section{Density of states\label{sect:DOS}}

In finite magnetic field, the dot energy levels are given by $\varepsilon
_{m\sigma }=\varepsilon _{d}+(\Delta /2+\mu _{v}B)(\delta _{m,e}-\delta
_{m,o})+g\mu _{B}B(\delta _{\sigma ,\uparrow }-\delta _{\sigma ,\downarrow
}) $, where $\varepsilon _{d}$ is the dot bare energy level and $B$ is the
applied magnetic field. $\Delta $ is the zero-field valley splitting and $%
\mu _{v}$ is the constant valley splitting slope. Note that $\Delta $ and $%
\mu _{v}$ constants are sample-dependent and may vary in actual experiments.
In the following, we have taken $\mu _{v}=0.1~\mathrm{meV/T}$, slightly
smaller than $g\mu _{B}=0.114~\mathrm{meV/T}$ where $g=2$. We have also
chosen the parameters $\Gamma =0.2$ meV, $D=200\Gamma $, and $U=40\Gamma $,
which, though bringing forth small Kondo temperatures, are favorable for the
purpose of illustration. 
\begin{figure}[tbp]
\vspace{-5 cm} \centering     
\subfigure[~DOS for $\mathcal{N}=1$
regime.] {\ \label{sekondo}
\includegraphics[width=0.47\textwidth]{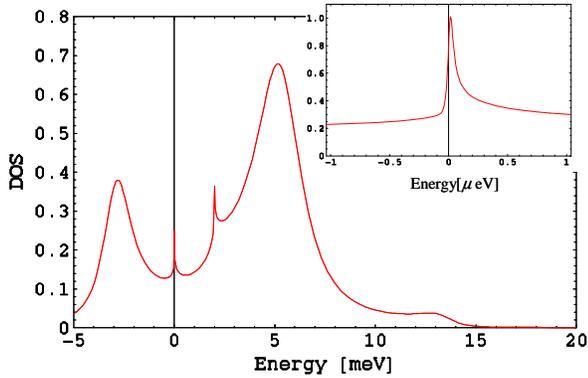}} \\ 
\vspace{-5 cm} 
\subfigure[~DOS for $\mathcal{N}=2$ regime.] {\ \label{dekondo}
\includegraphics[width=0.45\textwidth]{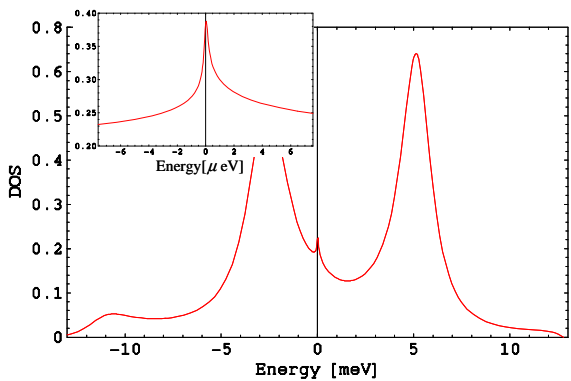} } \\ 
\vspace{-5 cm} 
\subfigure[~DOS for $\mathcal{N}=3$ regime.] {\ \label{tekondo}
\includegraphics[width=0.45\textwidth]{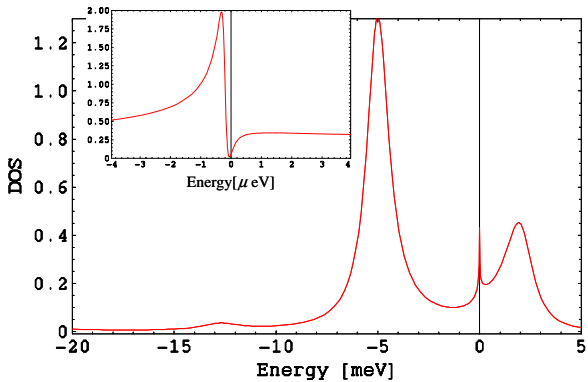} }
\caption{{\protect\small {\ The DOS's with approximately \protect\ref%
{sekondo} one, \protect\ref{dekondo} two, \protect\ref{tekondo} three
electrons on the dot. Here the valley and Zeeman splittings are assumed to
be zero. The valley index is considered conserved. In these three regimes
there is a Kondo resonance at the Fermi energy, shown in the insets. To plot
them we use \protect\ref{sekondo} $\protect\varepsilon _{d}=-15\Gamma $, $%
T/T_{K1}=0.2$ with $T_{K1}\sim 1\times 1^{-4}$ meV. \protect\ref{dekondo} $%
\protect\varepsilon _{d}=-55\Gamma $, $T/T_{K2}=0.2$ with $T_{K2}\sim
3\times 10^{-4}$ meV, \protect\ref{tekondo} $\protect\varepsilon %
_{d}=-105\Gamma $, $T/T_{K3}=0.4$ with $T_{K3}\sim 1.6\times 10^{-4}$ meV.
The self-consistently computed occupation numbers are \protect\ref{sekondo}, 
$<n_{m\protect\sigma }>=0.2505\approx 1/4;$ \protect\ref{dekondo}, $<n_{m%
\protect\sigma }>=0.4831\approx 1/2;$ \protect\ref{tekondo}, $<n_{m\protect%
\sigma }>=0.7312\approx 3/4$.}}}
\label{kondo123}
\end{figure}

\subsection{Degenerate spin-valley Kondo effect with valley index
conservation}

We have obtained in the previous section the $\beta $-dependent Kondo
temperatures. It is now instructive to apply them numerically as a
temperature scale to the Kondo resonance. Our task is to plot the DOS, each
time tuning the bare energy level $\varepsilon _{d}$ in order to shift the
Fermi energy $\varepsilon _{F}$ (which we set to be zero) inside the $%
\mathcal{N}=1,2,3$ Coulomb blockade regimes, mimicking experimental
manipulation of the gate voltage over the dot energy levels. Let us for a
moment consider valley index conservation and full spin-valley degeneracy by
disregarding the expected zero-field valley splitting. Figure \ref{kondo123}
displays three plots of the DOS near the Fermi energy in the $\mathcal{N}%
=1,2,3$ regimes. Not surprisingly, for odd $\mathcal{N}$ (1,3), a narrow
Kondo resonance at zero bias voltage can be seen in Fig.~\ref{sekondo}
and Fig.\ref{tekondo}. Another side peak around 2 meV in Fig.~\ref{sekondo}
comes from the process that two electrons are depleted from the dot
simultaneously and therefore is energetically disfavored at large $U$. It is
interesting to note that the positions and shapes of the Kondo resonances
for $\mathcal{N}=1,3$ reflect a particle-hole symmetry. Also noteworthy is
that in $\mathcal{N}=$ odd regimes the dot displays itself as a spin-1/2
magnetic impurity and is fully Kondo-screened by the conduction electrons on
the leads. The pseudospin possessed by the valley degree of freedom behaves
similarly.\

On the other hand, a more symmetrical zero-bias resonance in the $\mathcal{N}%
=2$ regime than the two in the $\mathcal{N}=1,3$ regimes is found in Fig.~%
\ref{dekondo}. This unexpected resonance is attributed to the valley degree
of freedom that provides valley states for additional inelastic transitions
to occur: for example, spin flips can occur even when the ground state is a
singlet, if the flip is accompanied by a change in valley index. In fact,
single-electron and double-electron Si QDs belong to the types of
fully-screened, and under-screened Kondo effect\cite{Bayani2007},
respectively. \ 

A somewhat similar phenomenon of the Kondo effect can occur in a two-level
quantum dot with orbital degeneracy \cite{Yeyati_1999}. By introducing an
interpolative perturbative approach the Kondo effect is obtained in the $%
\mathcal{N}=1,2,3$ regimes. However, this approach does not give an accurate
estimate of the Kondo temperatures as a function of model parameters and
only provides a qualitative thermal behavior. The EOM approach we have
chosen here acquires similar structures for the DOS and seems more favorable
to render better qualitative (if not quantitative) Kondo temperatures and
Kondo resonance structures.

\begin{figure}[htp]
\vspace{-6 cm} \centering      
\subfigure[\label{fig:Bfor2e1a}]  {
\includegraphics[width=0.5\textwidth ,clip]{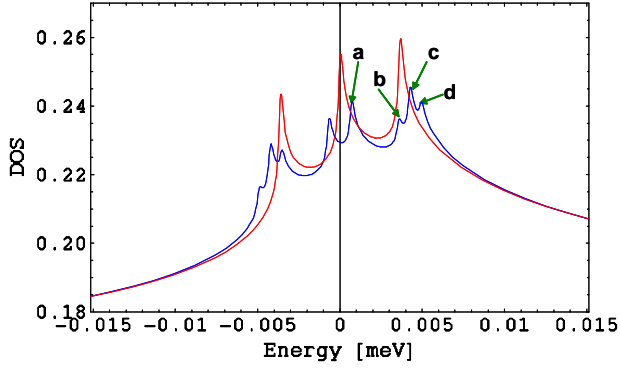}} \\  
\vspace{-2.5 in} \centering     
\subfigure[\label{fig:Bfor2e1b}]  {\hspace{1 cm}
\includegraphics[width=0.5\textwidth ,clip ]{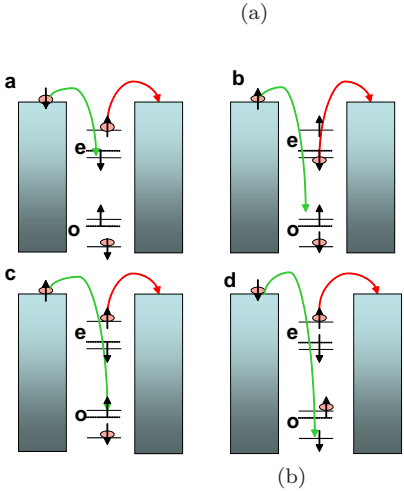}}
\caption{{\protect\small {(Color online) The DOS for $\mathcal{N}=2$ is plotted with $%
\protect\beta=0$ (valley index conservation), the zero-field valley
splitting $\Delta =12~T_{K2}$ and a magnetic field $B=0$ (red curve), $3\times 10^{-3}$~T (blue curve). The Kondo temperature $T_{K2}\sim 3\times 10^{-4}$
meV. $T/T_{K2}=0.2, ~\protect\varepsilon_d=-55\Gamma$, the same for Fig.~ 
\protect\ref{dekondo}. The split peaks and their field dependences parallel
those of single-electron Si QDs in Ref.~\protect\onlinecite{Shiau2007}, with similar
many-body transitions in the schematic diagrams $a,~b,~c,~d$ producing peaks 
$a,~b,~c,~d$, respectively. }}}
\label{fig:Bfor2e}
\end{figure}

\subsection{Field dependence and effect of valley index nonconservation for
the $\mathcal{N}=2$ regime}

The splitting of the Kondo resonance is instructive because it disentangles
clearly the interplay of all the participating low energy states and their
magnetic field dependences. The peak splittings are due to the lifting of
valley and spin degeneracies. Here we demonstrate the field-dependent peak
splittings while taking into consideration the effect of the valley index in
two extreme cases: conservation ($\beta =0$) and full nonconservation ($%
\beta =1$) of valley index. We shall particularly concentrate on the unusual
Kondo peak in the $\mathcal{N}=2$ regime.

\begin{figure}[tph]
\caption{{\protect\small {(Color online) The DOS for $\mathcal{N}=2$ is plotted with $%
\protect\beta =1$, the zero-field valley splitting $\Delta =10~T_{K2}^+(%
\protect\beta =1)$ and a magnetic field $B=0$ (red curve), $5\times 10^{-2}$~T (green curve). The Kondo temperature $T_{K2}^+(\protect%
\beta =1)\sim 5\times 10^{-3}$ meV. $T/T_{K2}^+(\protect\beta =1)=0.2,~%
\protect\varepsilon _{d}=-55\Gamma $. Note that there emerges an extra
zero-bias peak due to the effect of valley index nonconservation, compared
with the peak splitting structure in Fig.\protect\ref{fig:Bfor2e}, where the
valley index is conserved. It introduces another transition that involves an
electron hopping between opposite valleys, as shown in the schematic diagram.%
}}}
\label{fig:valleyB2e}\vspace{-7 cm} \centering{\ 
\includegraphics[width=0.55\textwidth
,clip]{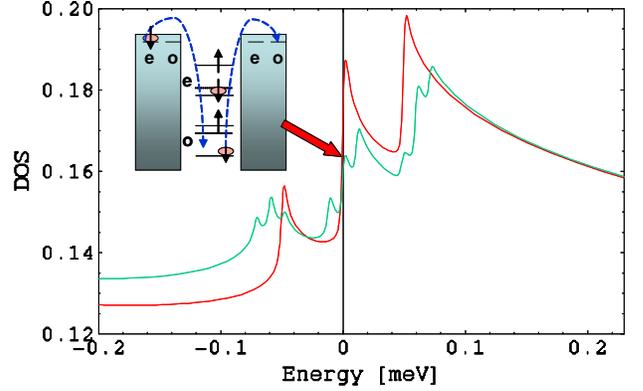}}
\end{figure}

Consider first $\beta =0$. Fig.~\ref{fig:Bfor2e} demonstrates the peak
splitting structure of double-electron Si QDs. The zero-bias peak splits
into three peaks in red curve when the zero-field valley splitting $\Delta
\not=0$ and the magnetic field $B=0$. Among these three peaks, the central
splits into two peaks by the Zeeman splitting when $B\not=0$, since it
originates from a spin Kondo effect, whereas each side peak splits further
into three; see the blue curve. Each peak correspond to a many-body
transition. \ The arguments parallel those for $\mathcal{N}=1$\cite%
{Shiau2007}: the spin-valley Kondo effect comes from spin-flip, intervalley,
and intravalley inelastic transitions among the dot states in Fig.~\ref%
{fig:config1a} interacting with the lead states. The valley and Zeeman
splittings break the degeneracy of these dot states to produce additional
peaks. \ We show the transition corresponding to each peak in the schematic
diagrams of Fig.~\ref{fig:Bfor2e1b}. \ 

At $\beta =1,$ we have maximal nonconservation of the valley index. \ For
this case, Figure \ref{fig:valleyB2e} shows an additional zero-bias peak
that represents a bound state of opposite valleys, which is therefore a a
manifestation of the pure \textit{valley} Kondo effect, apart from the other
eight peaks already identified in Fig.~\ref{fig:Bfor2e}. The transitions
that generate this peak take the following course: a conduction electron at
the odd (even) dot state tunnels out to the odd (even) lead state through
the intravalley coupling $V_{O}$, while another electron in the even (odd)
lead state tunnels into the odd (even) dot state through the intervalley
coupling $V_{X}$ ; see the schematic in Fig.~\ref{fig:valleyB2e}. This peak
height increases along with the intervalley coupling $V_{X}$, or $\beta $,
which would therefore provide an experimental signature of valley index
nonconservation.\

As already seen in Fig.~\ref{kondo123}, the $\mathcal{N}=1$ and $\mathcal{N}%
=3$ cases are rather similar. \ This is due to particle-hole symmetry:
instead of electrons, it is the holes that tunnel in and out of the dot. \
The four spin-valley energy states interacting with the lead states for $%
\mathcal{N}=3 $ case are shown in Fig.~\ref{fig:config1b}. With these four
states gradually separated by the valley and Zeeman splittings, the
inelastic co-tunnelings produce a similar peak splitting structure to that
for $\mathcal{N}=1$, a case which has been exhaustively treated in Ref.~%
\onlinecite{Shiau2007}. To avoid redundancy, we omit the plot of its DOS. \ 

\section{Conclusions}

We summarize our results as follows:(a) The spin-valley Kondo effect appears
as expected in the $\mathcal{N}=1$ and $\mathcal{N}=3$ regimes, and
unexpectedly also in the $\mathcal{N}=2$ regime, but not in the $\mathcal{N}%
=4$ regime, thus yielding in general a Kondo effect unless $\mathcal{N}$ is
divisible by 4. \ This contrasts with the spin Kondo effect which appears
only at odd $\mathcal{N}.$ \ (b) Figure~\ref{kondo123} shows an asymmetrical
structure astride the Fermi energy in the position and shape of the
zero-bias Kondo peaks in the $\mathcal{N}=1$ and $\mathcal{N}=3$ regimes. In
the $\mathcal{N}=2$ regime the peak shape is more symmetrical. (c) By
applying a magnetic field, the peak splittings in the $\mathcal{N}=2$
regimes resemble that in the $\mathcal{N}=1$ regime. We are able to to
attribute each peak to its corresponding inelastic many-body transition
(co-tunneling); see the cartoon schematics in Fig.~\ref{fig:Bfor2e}. \ The
rich level structure gives rise to a rich pattern of peaks.

We expect that these many-body signatures of the valley degree of freedom in
Si will be observed in future experiments.

\begin{acknowledgments}
We would like to thank S. Chutia, L.J. Klein, M. Friesen and M.A. Eriksson
for useful discussions. \ This work was supported by NSA and ARDA under ARO
contract number W911NF-04-1-0389 and by the National Science Foundation
through the ITR (DMR-0325634) and EMT (CCF-0523675) programs.
\end{acknowledgments}

\end{document}